\documentclass[conference]{IEEEtran}

\newlength\figwidth
\IEEEoverridecommandlockouts
\usepackage{cite}
\usepackage{amsmath,amssymb,amsfonts}
\usepackage{algorithmic}
\usepackage{graphicx}
\graphicspath{{Figures_pdf/}}

\usepackage{textcomp}
\usepackage{xcolor}
\usepackage{url}
\usepackage[bookmarks=false]{hyperref}
\hypersetup{
	linktocpage=true,
	colorlinks,
	urlcolor=blue,
	linktocpage=true,
	pdfborderstyle={/S/S/W 1},
	hyperindex=true,
	bookmarks=true,
	bookmarksopen=true,
	bookmarksnumbered=true,
}

\begin{document}
	
	\title{Network Analysis of Chaotic Dynamics in Fixed-precision Digital Domain
		\thanks{This work was supported by the National Natural Science Foundation of China (nos. 61772447, 61532020).}
	}
	\author{\IEEEauthorblockN{Chengqing Li}
		\IEEEauthorblockA{\textit{College of Computer Science and}\\
			\textit{Electronic Engineering,} \textit{Hunan University,}\\
			 Changsha 410082, Hunan, China\\
			\href{chengqingg@gmail.com}{\nolinkurl{chengqingg@gmail.com}} }
		\and
		\IEEEauthorblockN{Jinhu Lu}
		\IEEEauthorblockA{\textit{School of Automation Science}\\
			\textit{and Electrical Engineering,} \\
			\textit{Beihang University,}\\
			Beijing 100083, China}
		\and			
		\IEEEauthorblockN{Guanrong Chen
		\IEEEauthorblockA{\textit{Department of Electronic Engineering,} \\
			\textit{City University of Hong Kong, }\\
		Hong Kong SAR, China}}	
	}
	
\maketitle

\begin{abstract}
When implemented in the digital domain with time, space and value discretized in the binary form,
many good dynamical properties of chaotic systems in continuous domain may be degraded or even
diminish. To measure the dynamic complexity of a digital chaotic system, the dynamics can be
transformed to the form of a state-mapping network. Then, the parameters of the network are
verified by some typical dynamical metrics of the original chaotic system in infinite precision,
such as Lyapunov exponent and entropy. This article reviews some representative works on the
network-based analysis of digital chaotic dynamics and presents a general framework for such
analysis, unveiling some intrinsic relationships between digital chaos and complex networks.
As an example for discussion, the dynamics of a state-mapping network of the Logistic
map in a fixed-precision computer is analyzed and discussed.
\end{abstract}


\section{Introduction}

\IEEEPARstart{C}haos is typically defined in the continuous domain, for which many interesting
dynamical properties hold only in infinite precision. Yet, implementing a chaotic map in a
finite-precision device is an inevitable process for any of its real applications or for any
simulation in the digital world. Therefore, the dynamical properties of a digitized chaotic map
have been the main concern in nonlinear science and chaotic cryptography. When iterating a chaotic map in the digital domain, its orbit is perturbed by the quantization errors. Consequently, not a real-valued chaotic
orbit but only a discretized pseudo-orbit is obtained. Such pseudo-orbits are considered useful
due to the well-known \textit{shadowing lemma}, which says that every pseudo-orbit obtained
via computer simulation has a true error-free orbit nearby. The shadowing lemma justifies the
use of computers to study chaotic systems and their dynamics. As a consequence, many efforts
were devoted to designing pseudo-random number generators utilizing chaotic maps
\cite{Phatak:LogisticRNG:PRE95}. Nevertheless, it was argued
\cite{Persohn:AnalyzeLogistic:CSF12,cqli:network:TCASI2019} that the dynamics of digital chaotic
maps are degraded to various extents and such dynamical degradation can create big problems
to some applications, for example thwarting the security of the chaos-based cryptosystems 
as iterating a degraded chaotic map generates a short-period pseudo-random number sequence with a 
non-negligible probability \cite{cqli:autoblock:IEEEM18,cqli:IEAIE:IE18,Yoshioka:ChebyshevPk:TCAS2:2018}.

Despite the theoretic elegancy of the shadowing lemma, it becomes rather misleading when one
looks at the system as a whole: all shadowing orbits expressed in a particular finite precision
constitute only a negligible part of the whole set of true orbits. This is not surprising because
the states that a pseudo-orbit can visit are countable, but in continuous domain they are not.
Even worse is that, in a finite-state space, all pseudo-orbits will eventually become periodic
and the period is bounded by the number of the states therein. Thus, to what extent the
finite-precision orbits can still retain the dynamical properties of the original chaotic system
calls for quantitative analysis of the set of pseudo-orbits. This turns out to be a rather
complicated problem and cannot be easily handled by existing theories and techniques. In fact,
most existing work focus on the average behaviors of all pseudo-orbits, but still cannot tell
anything more about the pseudo-orbits within different fine structures. In \cite{SJLi:degradation:IJBC05},
some fine structures in such pseudo-orbits implemented with fixed-point arithmetic are revealed
and discussed.

Recently, complex networks as an emerging research field can provide useful tools for
studying the dynamics of chaotic time series, first attempted in \cite{Zhang:NetworkTimeSeries:PRL2006}.
Essentially, some earlier work studying the dynamics of digital chaos can be attributed to the
notion of complex networks \cite{zouyong:network:PR19}. Specifically, in 1986 Binder and Jensen built
a state-mapping network (SMN) of the Logistic map and reported that the counterparts of some metrics on
measuring the dynamics of continuous chaos, such as Lyapunov exponent, remain to work as usual
\cite{Binder:Logistic:PRA86}. In the following years, Binder reported some more work
and findings: the relationship between a limit cycle of the Logistic map and its control parameter
\cite{Binder:cycles:PHyD1992}. Binder and Okamoto studied the shadowing of unstable periodic orbits
(UPO) with the corresponding limit cycles \cite{Binder:periodicorbits:2003}. In \cite{Hsu:cell:IJBC92},
Hsu investigated the global properties of nonlinear dynamical systems by cell-to-cell mapping.
Kurths and his team presented an analytical framework for recurrence network analysis of time
series \cite{Kurths:SeriesNetwork:PRE2012}, and then further studied the geometry of chaotic
dynamics from the perspective of complex networks \cite{Donner:GeometryDynamics:EPJB2011}.
In \cite{Frahm:UlamNetwork:PRE18}, Frahm and Shepelyansky analyzed the small-world properties of
Ulam's networks of the Chirikov standard map and the Arnold cat map, showing that the number of
degrees of separation grows logarithmically with the network size. In \cite{Shreim:NetworkCA:2007},
Shreim et al. studied the dynamics of series generated by cellular automata, using the parameters
of the associate state-mapping network. The properties of Chebyshev map on a finite field was
investigated with a directed network (graph) by Gassert \cite{Gassert:Chebyshev:DM2014}. Recently,
the \textit{functional graph} associated with the iterations of the Chebyshev polynomials over a finite field
was quantitatively analyzed by Qureshi and Panario \cite{Qureshi:Chebyshev2018}. It should be noted, however, that networks typically exhibit self-organizing fractals
(and multifractals) and chaos by themselves when they are sufficiently large and complex
\cite{song:fractal:nature05}.

This paper reviews the existing complex-network-based dynamic analysis of digital chaotic systems,
using the Logistic map as a typical example. A general framework for analysis is presented.
Furthermore, it discusses how the structure of the SMN of the Logistic map is changed with the
fixed-point arithmetic precision on a computer. Some fractal phenomena and power-law in-degree
distributions are observed when the precision is incrementally increased.

The rest of the paper is organized as follows. Section~\ref{survey} briefly reviews the existing
dynamic analysis of digital chaotic systems from the general viewpoint of complex networks. Some
special properties of the SMN of the Logistic map are reported in Sec.~\ref{Dynamic}.
The last section concludes the paper.

\section{Brief survey on complex-network-based dynamic analysis of digital chaotic maps}
\label{survey}

Following Ulam's method proposed in \cite[p. 74]{Ulam:Math:1964}, the state-mapping network
of the Logistic map
\begin{equation}
f(x)=\mu \cdot x \cdot (1-x)
\label{Logistic}
\end{equation}
was built by the following three steps in \cite{Iba:NetworkChaos:ICCS2011,Thurner:Logistic:ICCS2007}:
1) divide the phase interval into some subintervals;
2) each subinterval is considered as a node;
3) two nodes are linked by a directed edge if a (predefined) map relationship between the corresponding
intervals exist.

\begin{figure}[!htb]
\center
\includegraphics[width=1.4\figwidth]{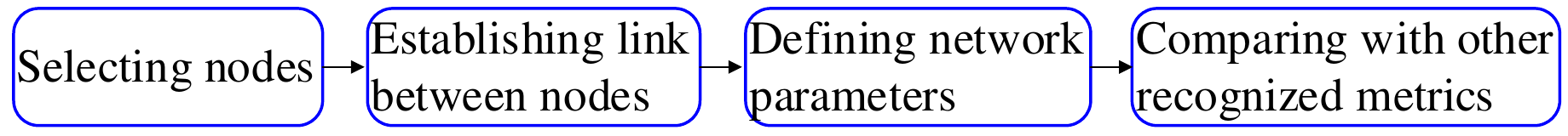}
\caption{Framework of network-based analysis of dynamical systems.}
\label{figure:Framework}
\end{figure}

Once the network is built, various parameters can be calculated to reveal some rules
similar to that obtained in the infinite resolution domain. The general framework of analyzing
the dynamical properties of chaotic systems can be shown by the four stages in
Fig.~\ref{figure:Framework}.

In \cite{Thurner:Logistic:ICCS2007}, the phase space of a chaotic map $g(x)$ is equally divided
into $N$ disjoint sub-space, and each sub-space is defined as a node. Given an initial state,
$x_0$, the $i$th node is directly linked to the $j$th node if and only if $x_0$ and $g(x_0)$ belong to the
two corresponding sub-spaces, respectively. Iterating the chaotic system from the current state
$T$ several times and linking the related nodes in a similar manner, the space network is established.

\begin{figure}[!htb]
	\center
\includegraphics[width=\figwidth]{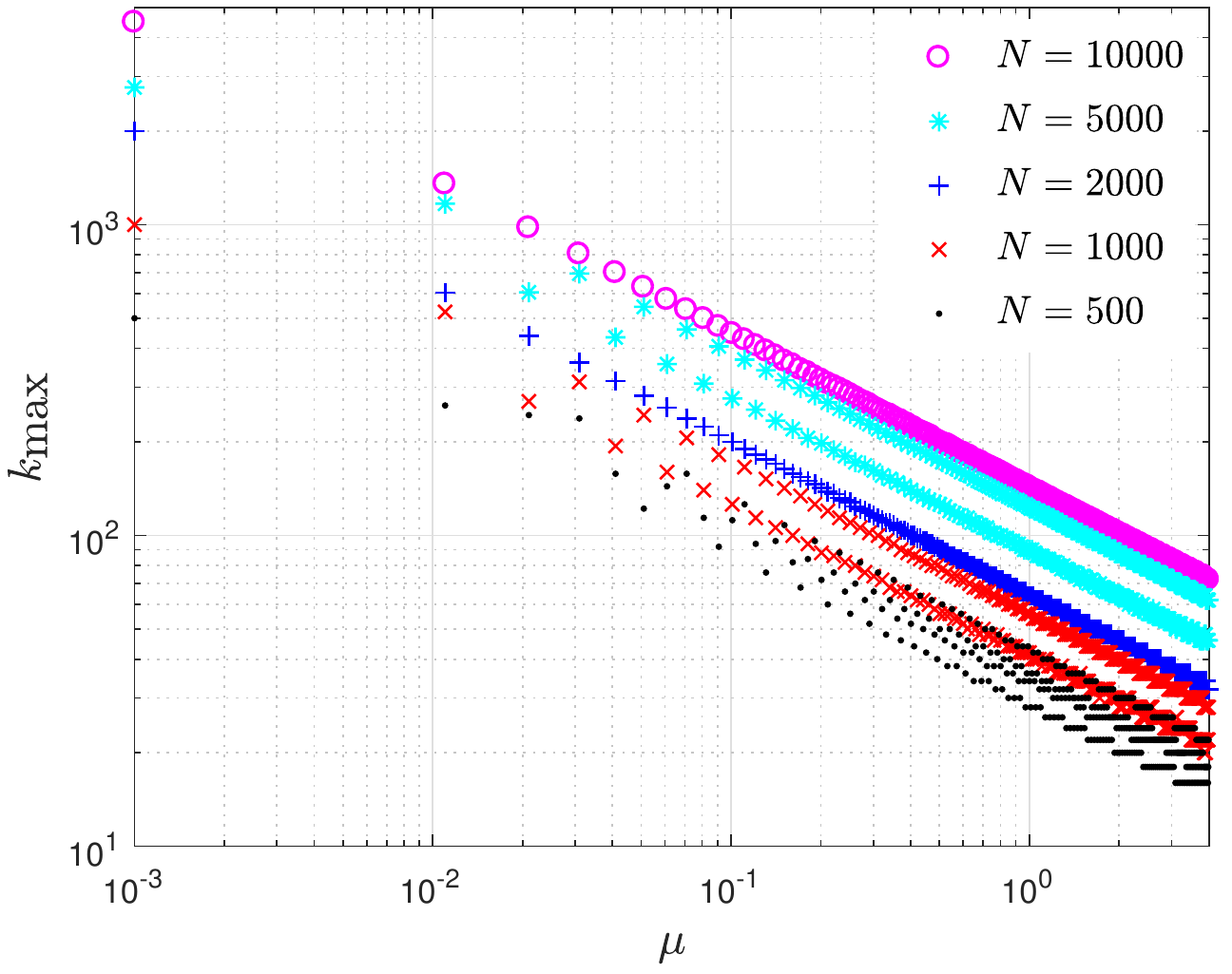}
	\caption{The largest in-degree $k_{\mbox{max}}$ of the space network of the Logistic map
in five network sizes.}
	\label{fig:logisticdistribution}
\end{figure}

Figure~\ref{fig:logisticdistribution} depicts the largest in-degree of the corresponding space
network under different control parameters with five network sizes.
From Fig.~\ref{fig:logisticdistribution}, one can see that the largest in-degree obviously follows
a power-law distribution with the same scaling exponent for $N\in\{2000, 10000\}$. In contract,
there are two scaling exponents for $N\in\{1000, 5000\}$. This may be caused by the non-uniform
distributions of the floating-point numbers in different sub-spaces.

\begin{figure}[!htb]
	\center
\includegraphics[width=\figwidth]{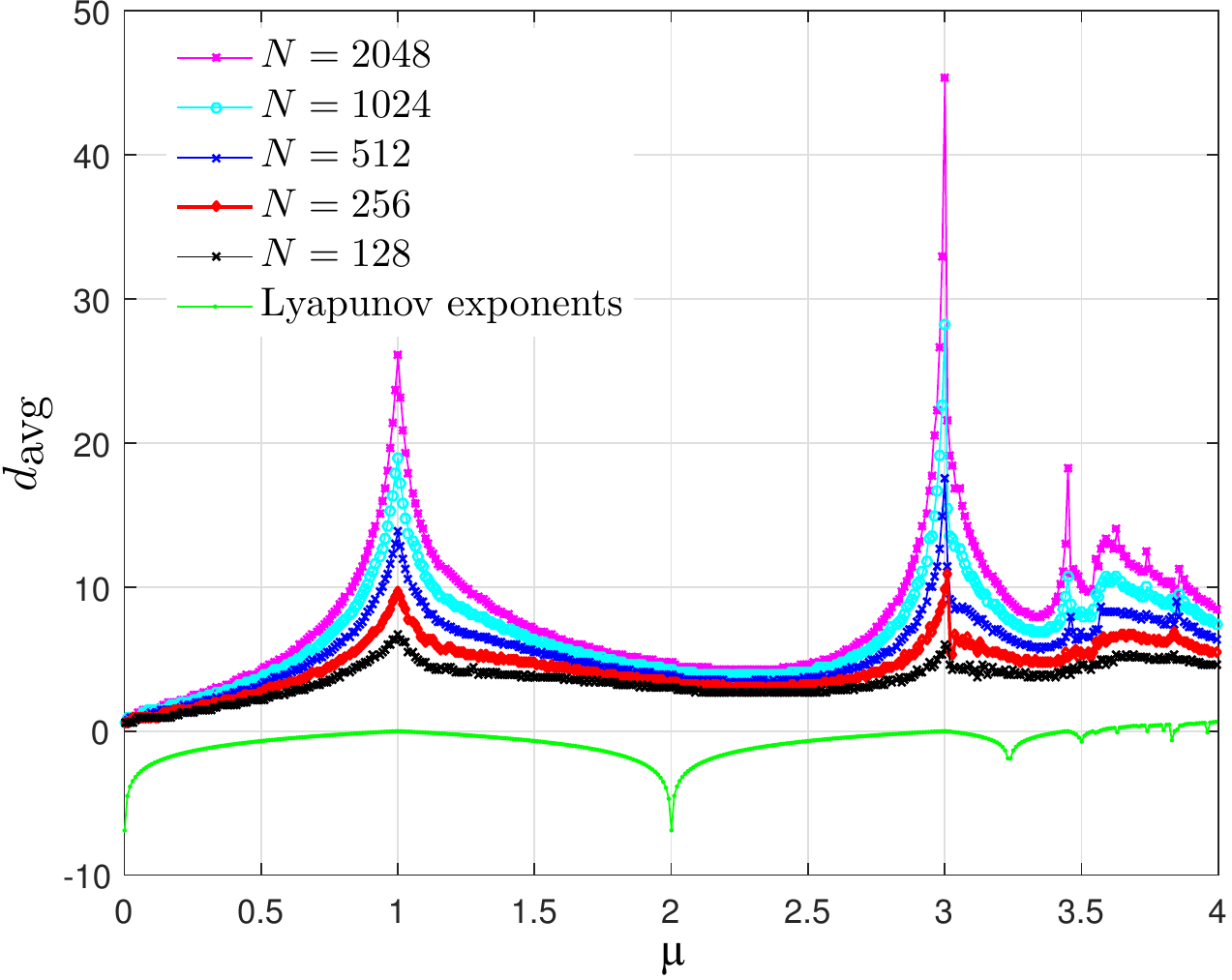}
	\caption{The average path length of the space network of the Logistic map with respect to
the control parameter $\mu$ for five network sizes.}
	\label{fig:logisticdistance}
\end{figure}

Figure~\ref{fig:logisticdistance} presents the average path length of the space network of the
Logistic map under different control parameters with five network sizes. Observing
Fig.~\ref{fig:logisticdistance}, one can see that the changing trend of the average path length
with respect to the control parameter matches that of the Lyapunov exponents of the Logistic map
to a relatively high extent, demonstrating that the parameters of the phase network can serve as
an alternative measure for detecting the `edge of chaos' in dynamical systems.

\begin{figure}[!htb]
\center\includegraphics[width=\figwidth]{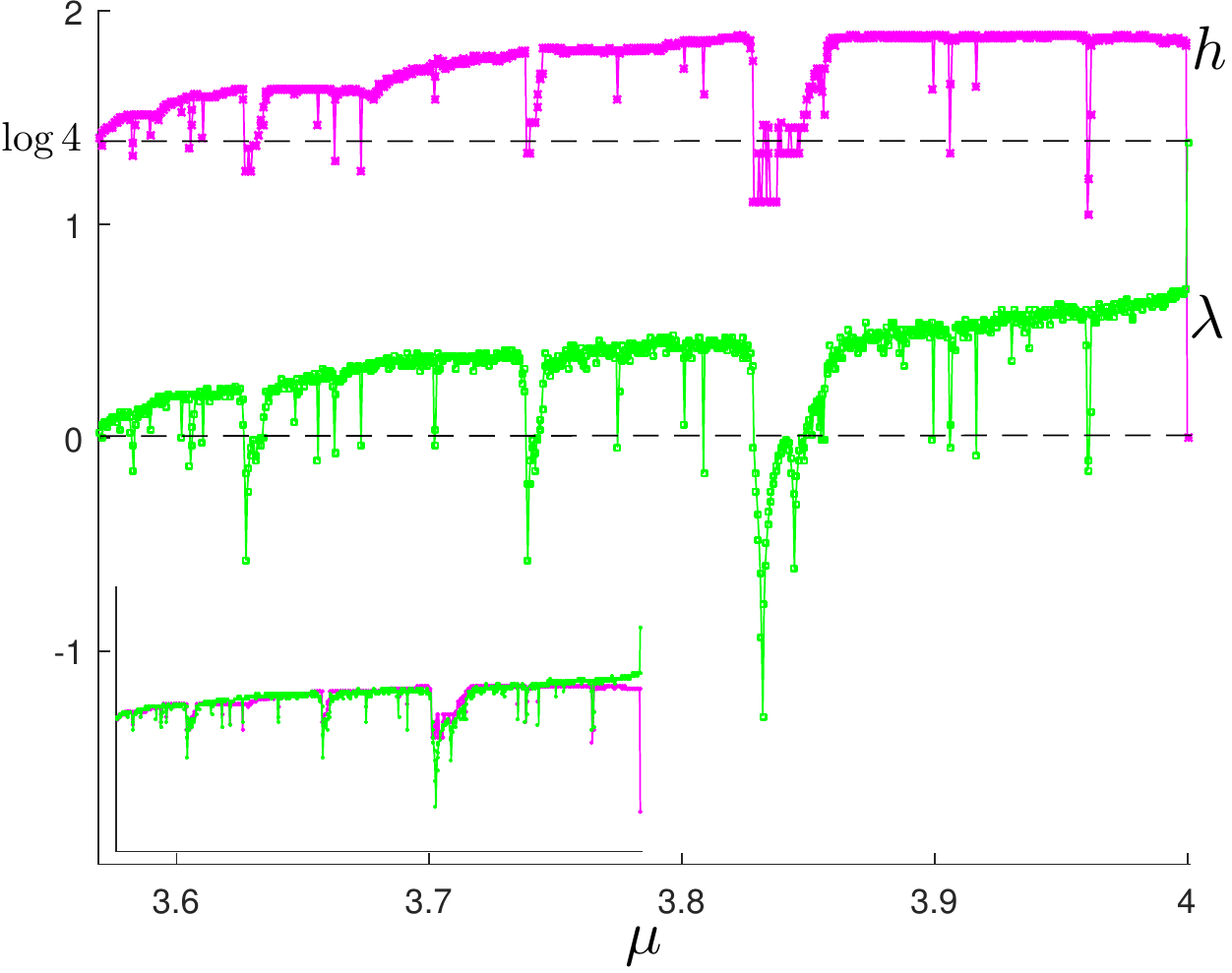}
\caption{Dynamical demonstration of the Logistic map under two metrics: horizontal visibility
		network entropy $h$ and Lyapunov exponent $\lambda$.}
\label{figure:feigenbaum}
\end{figure}

\begin{figure}[!htb]
\center
\includegraphics[width=\figwidth]{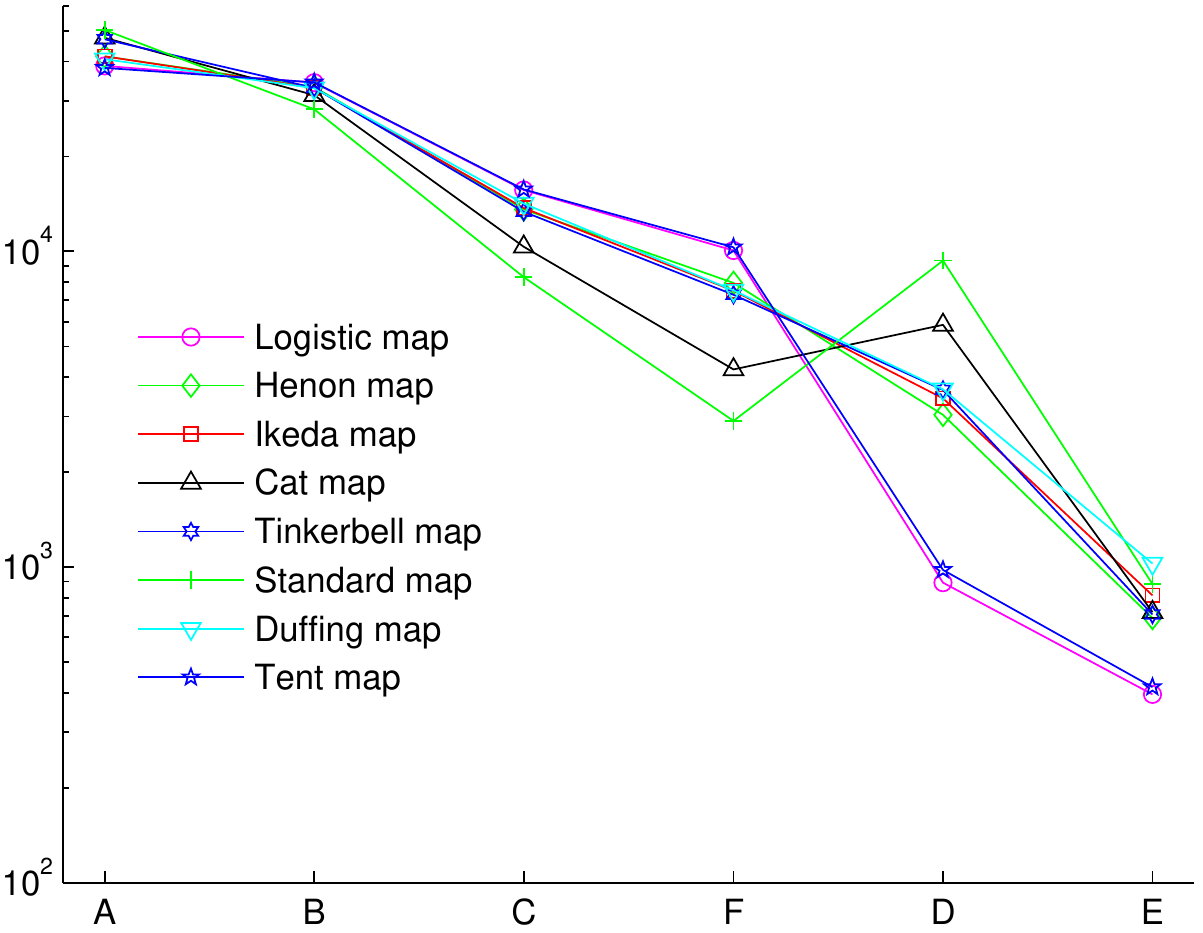}
\caption{Subgraph ranks of different types of time series, where $A$, $B$, $C$, $D$, $E$, and $F$
represent 6 different possible undirected subgraphs of size 4.}
\label{fig:motif}
\end{figure}

In \cite{Luque:FeigenbaumChaos:PLOS11}, the time series generated by a chaotic system is transformed
to a network (graph) using the mechanism of \textit{horizontal visibility graph}, where every state is
considered as a node and two arbitrary data points are linked if and only if they satisfy the geometric
`visibility' criterion. Then, a parameter called ``network entropy" is defined. As shown in
Fig.~\ref{figure:feigenbaum}, the changing trend of the network entropy of the associate network of the
Logistic map can mimic that of the Lyapunov exponent. In \cite{xu:motif:PNAS08}, two arbitrary states
of a time series are connected as a pair of nodes if the linear interpolation between them is larger
than the values of all intermediate points. The distribution of different subgraphs is shown to be a
sensitive measure of the underlying dynamics. As shown in Fig.~\ref{fig:motif}, motif distribution
can be used to distinguish broad classes of chaotic dynamical systems, to which the Logistic map belongs.

\begin{figure}[!htb]
	\centering
	\includegraphics[width=\figwidth]{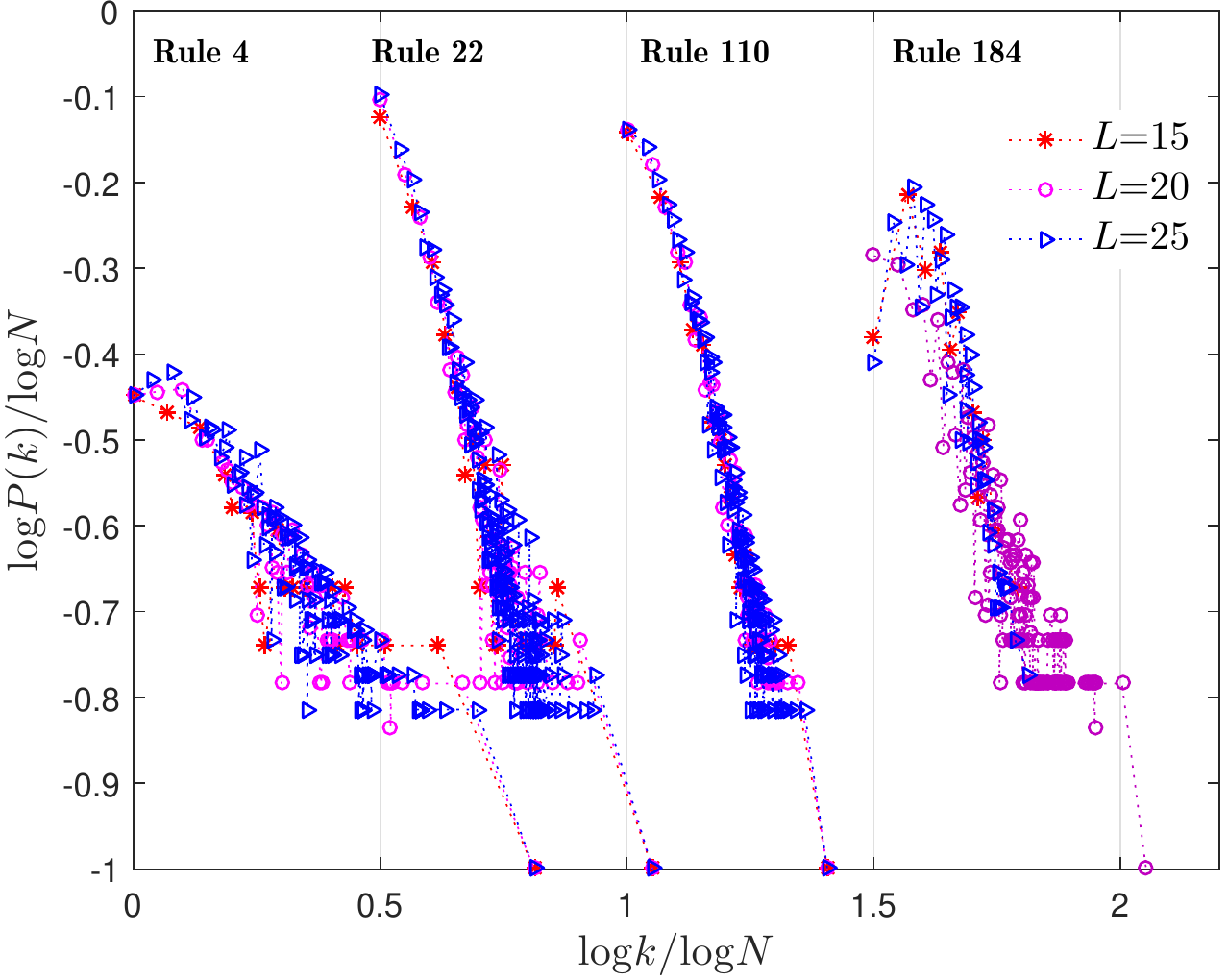}
	\caption{In-degree distribution $P(k)$ for four rules of CA with three values of CA having size
		$L$, where $N=2^L$.}
	\label{fig:IndegreeDistribution_CA}
\end{figure}

In \cite{Shreim:NetworkCA:2007}, the state-space network of the \textit{sequential dynamical systems}
is analyzed, which is composed by 1-D cellular automata (CA) with rules involving two colors and the
nearest neighbors, where every binary state of size $L$ is viewed as a node of a directed network and
the evolution relation between two nodes builds a link connecting them. Based on analytic analyses
and experimental results on some selected rules, it was claimed that the co-appearance of nontrivial
scalings in both the hub sizes and the path diversity of the state-phase networks can separate simple
dynamics from the more complex ones found in CAs falling into Wolfram's classes III and IV (see the
in-degree distribution for four rules shown in Fig.~\ref{fig:IndegreeDistribution_CA}). Yet, the
validity of such classification method was questioned in \cite{Cqli:CA:IJBC2017}, which plots the
distribution of the two parameters corresponding to all elementary rules of 1-D CA falling into the
same Wolfram's class. How to accurately measure the dynamical complexity of a chaotic system with
the methodology of complex networks needs further investigation.

It has been observed that, in the field of nonlinear science, many did not care about how their
computers perform fixed-point or floating-point arithmetic operations, but implemented a chaotic
system in the computer just like a black box. Most related research on dynamical analysis based on
network (graph) adopt the Logistic map as test objects \cite{Binder:Logistic:PRA86,
Phatak:LogisticRNG:PRE95,Thurner:Logistic:ICCS2007,xu:motif:PNAS08,Iba:NetworkChaos:ICCS2011,
Donner:GeometryDynamics:EPJB2011,Luque:FeigenbaumChaos:PLOS11,Persohn:AnalyzeLogistic:CSF12,
Kurths:SeriesNetwork:PRE2012}, where the real structure of SMN of the Logistic map in the computer
has been unfortunately ignored.

\section{Dynamical analysis of a state-mapping network of the Logistic map on computer}
\label{Dynamic}

Now, consider the following \textit{state-mapping network} of the Logistic map: every value in the
$n$-bit arithmetic computing precision is identified as a node; a directed edge between nodes $u$
and $v$ exists if and only if the former node is mapped to the latter one by the Logistic map.
The following key parameters of the space network were studied, revealing the relationships among
the Logistic map~\eqref{Logistic}, the computing precision and the control parameter.
\begin{itemize}
	\item \textit{Loop}: (also called a self-loop) is an edge that connects a node to itself.
	\item A directed \textit{cycle} in a directed graph is a sequence of nodes, starting and ending
at the same node such that, for each pair of two consecutive nodes in the cycle, there exists an
edge directed from the previous node to the later one.
	\item A \textit{weakly connected component (WCC)} is a maximal subgraph of a directed graph such that
	there is a path between every pair of nodes $u$ and $v$ in the subgraph, regardless of the direction of edges.
\end{itemize}

\begin{figure}[!htb]
	\centering
	\begin{minipage}{0.47\figwidth}
		\centering
		\includegraphics[width=\textwidth]{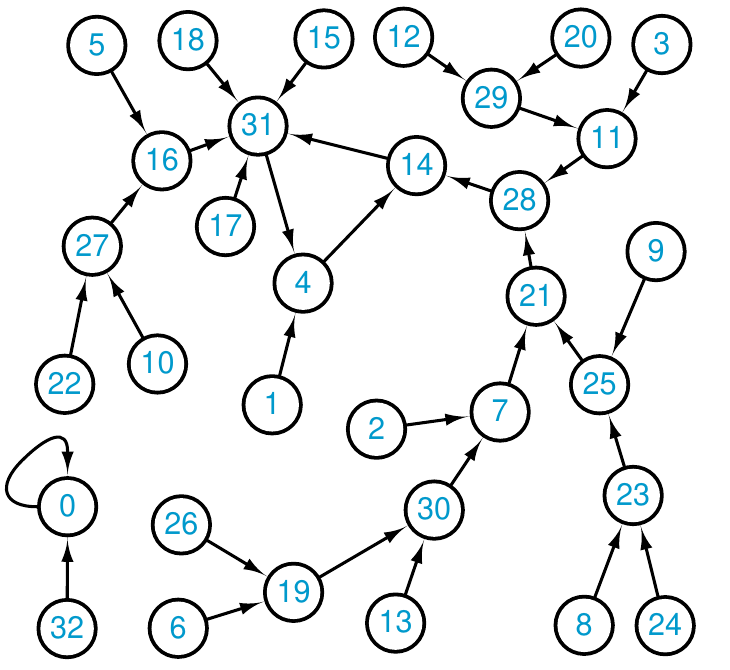}
		a)
	\end{minipage}
	\begin{minipage}{0.53\figwidth}
		\centering
		\includegraphics[width=\textwidth]{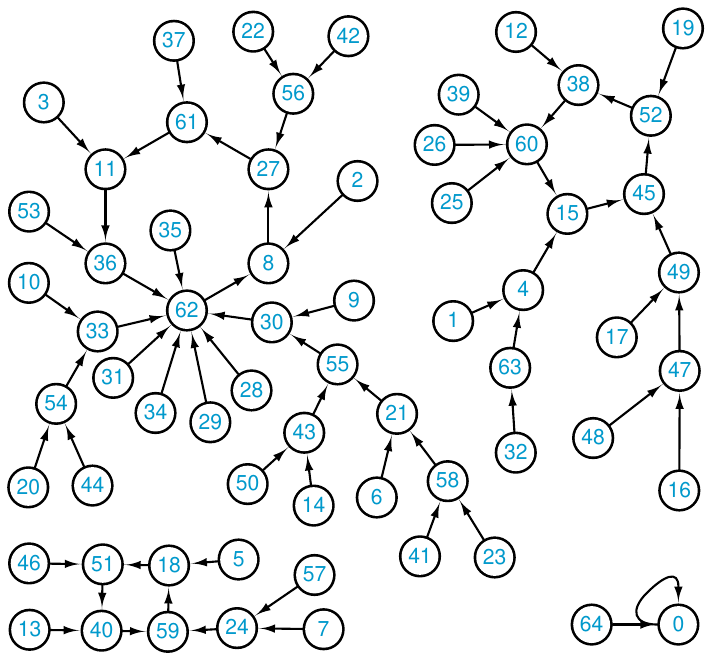}
		b)
	\end{minipage}
	\caption{The state network of the Logistic map with $\mu=\frac{121}{2^5}$ under round quantization:
		a) 5-bit precision; b) 6-bit precision.}
	\label{fig:networkLogistic5and6bits}
\end{figure}

\begin{figure}[!htb]
	\centering
	\includegraphics[width=0.85\figwidth]{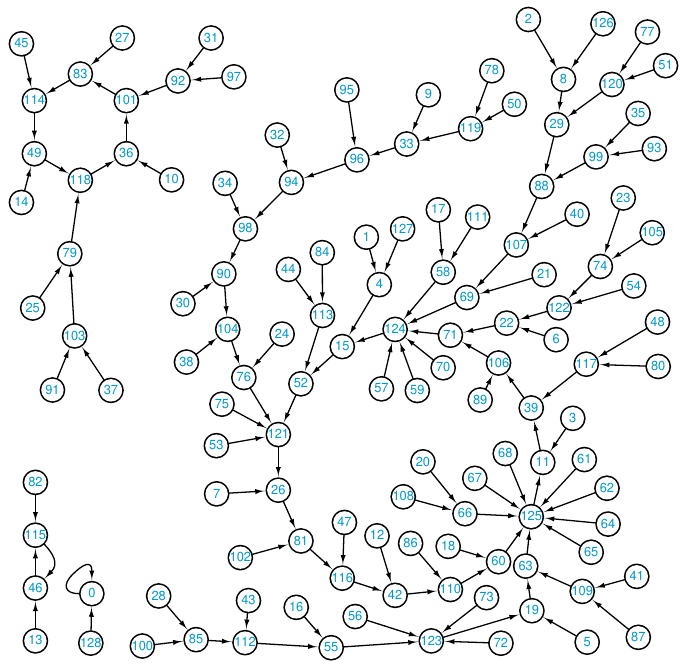}
	\caption{The state network of the Logistic map under 6-bit precision and round quantization, where
		$\mu=\frac{484}{2^7}$.}
	\label{fig:networkLogistic6bits}
\end{figure}

\begin{figure}[!htb]
	\centering
	\includegraphics[width=1.1\figwidth]{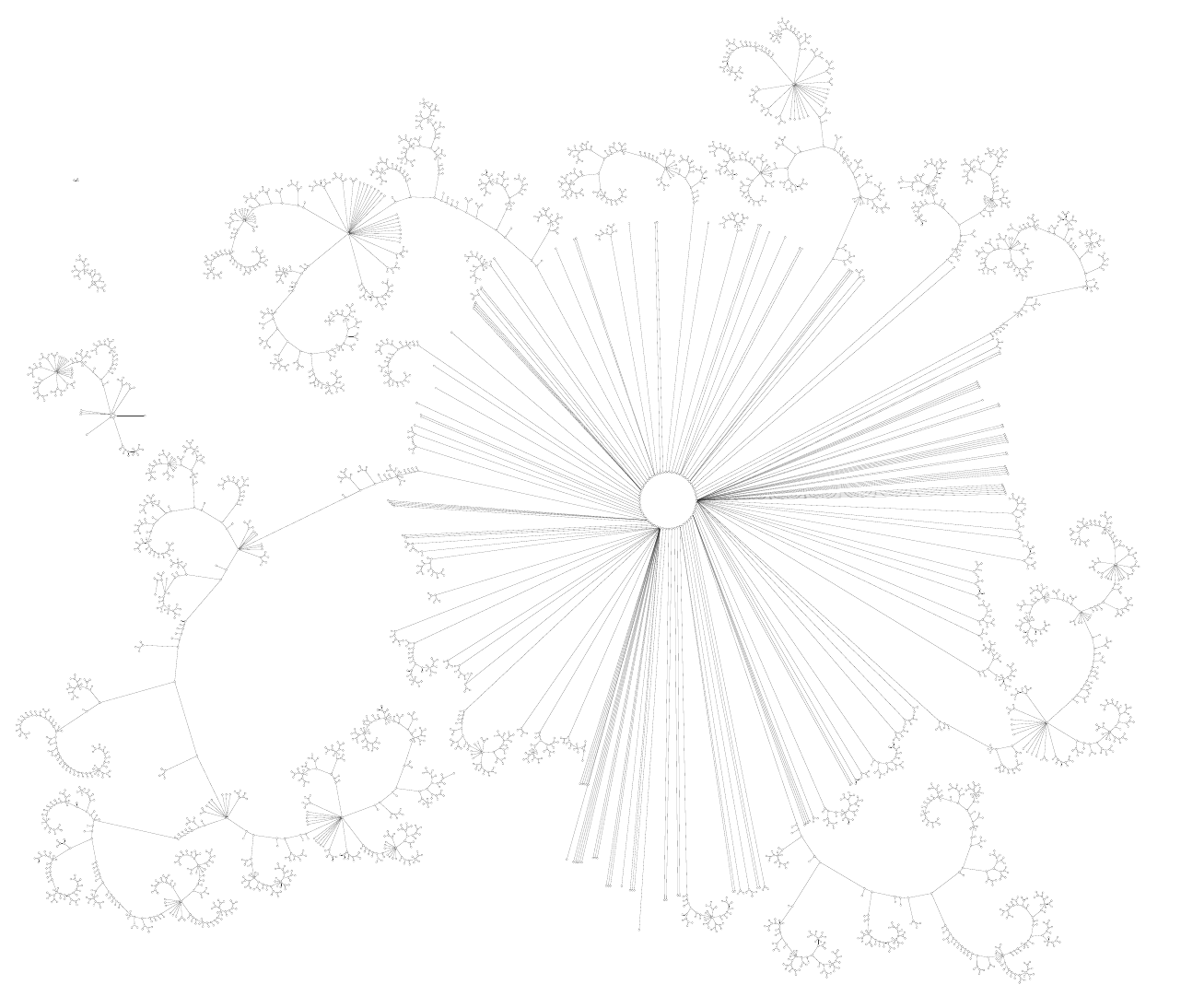}
	\caption{The state network of the Logistic map under 12-bit precision and round quantization, where
		$\mu=\frac{15488}{2^{12}}$.}
	\label{fig:statenetwork12bit}
\end{figure}

\begin{figure}[!htb]
	\centering
	\includegraphics[width=\figwidth]{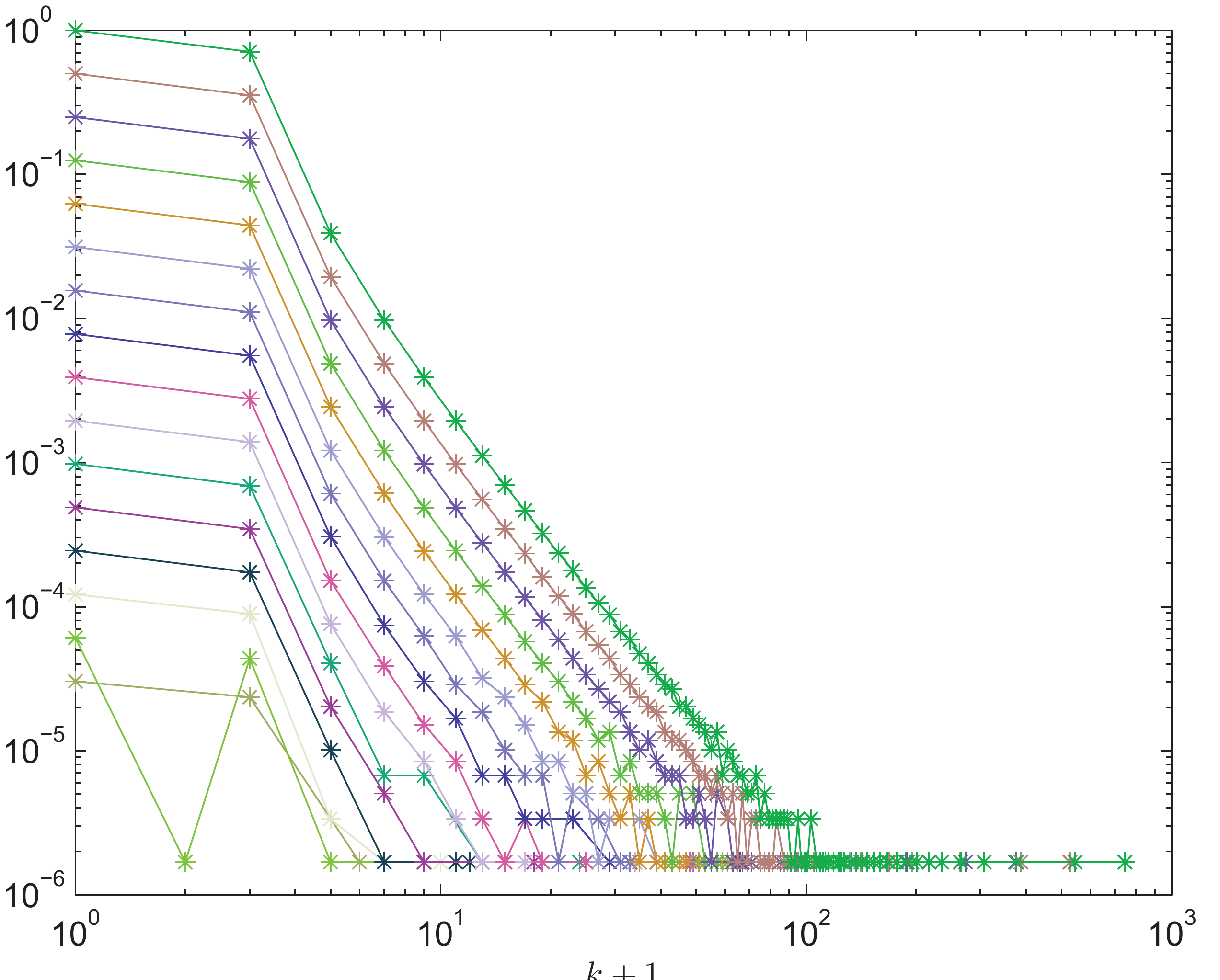}
	\caption{In-degree distribution of the sate networks of $F_{5} \sim F_{20}$.}
	\label{fig:InDegreeDistribution5_20}
\end{figure}

To observe the properties of SMN of the Logistic map, the software platform Cytoscape is used to
draw some space networks under some given parameter values. The space networks of the Logistic map
of $\mu=\frac{121}{2^5}$ under 5-bit and 6-bit fixed-point arithmetic are shown in
Fig.~\ref{fig:networkLogistic5and6bits}a) and b), respectively, where a round quantization is
adopted. To show the rule of the network with respect to the increase of the precision more clearly,
the state network of the Logistic map under 6-bit precision with the same value of the control
parameter is shown in Fig.~\ref{fig:networkLogistic6bits}. When the precision is further increased
to 12, the network becomes the case shown in Fig.~\ref{fig:statenetwork12bit}, demonstrating some
fractal phenomena, as studied in \cite{ZYu:FractalBrownianNetwork:PRE14}. As shown in
Fig.~\ref{fig:InDegreeDistribution5_20}, the in-degree distribution of $F_{n}$ approaches a
power-law more steadily as $n$ increases, where $F_{n}$ denotes the function of Logistic map~\eqref{Logistic}
implemented under $n$-bit precision.

\begin{table*}[!htb]
	\centering
	\caption{Intermediate values of the calculating Logistic map in binary16.}
	\begin{tabular}{c *{4}{|c}}
		\hline
		$x$    &  $1-x$ & $1-(1-x)$  &   $f(x)$  &   $f(1-x)$ \\ \hline
		0.0099945068359375 & 0.98974609375 & 0.01025390625 & 0.037384033203125 & 0.038360595703125\\ \hline
		0.04998779296875 & 0.94970703125 & 0.05029296875 & 0.179443359375 & 0.1805419921875\\ \hline
		0.0899658203125  & 0.90966796875 & 0.09033203125 & 0.309326171875 & 0.310546875\\ \hline
		0.0999755859375  & 0.89990234375 & 0.10009765625 & 0.340087890625 & 0.340576171875\\ \hline
		0.199951171875   & 0.7998046875 & 0.2001953125 & 0.6044921875 & 0.60498046875\\ \hline
		0.289794921875   & 0.7099609375 & 0.2900390625 & 0.77783203125 & 0.7783203125\\ \hline
		0.389892578125   & 0.60986328125 & 0.39013671875 & 0.89892578125 & 0.8994140625\\ \hline
		0.489990234375   & 0.509765625 & 0.490234375 & 0.9443359375 & 0.94482421875\\ \hline
	\end{tabular}
	\label{tab:differenceLogistic}
\end{table*}

Based on the above-verified results, one could infer more on the properties of the space network of the
Logistic map, as listed below.
\begin{itemize}
	\item Each space network is composed of a small number of weakly connected components;
	\item Each weakly connected component has one and only one self-loop or cycle;
	\item In each cycle of length longer than 2, there is at least one node with a large in-degree;
	\item Let $(D_n/2^n)$ denote the node with the largest degree in the state network of $F_n(x)$.
Then,
	\begin{equation}
	D_{n+1}\in \{2\cdot D_n, 2\cdot D_n+1, 2\cdot D_n-1\},
	\end{equation}
	where $n\ge n_\mu$, $\mu=\frac{N_\mu}{2^{n_\mu}}$, $N_\mu,  n_\mu\in \mathbb{Z}^+$.
	\item One weakly-connected component dominates the whole network, i.e., the size of the component
accounts for more than half of the size of the whole network. Among all the connected components,
there is a clearly decreasing order.
\end{itemize}

In the floating-point arithmetic domain, the symmetric property of the Logistic map does not hold in
general \cite{cqli:network:TCASI2019}. To demonstrate this point, some intermediate data obtained from calculating the Logistic map using half-precision floating-point format (binary16) are shown in Table~\ref{tab:differenceLogistic}.

\section{Conclusion}

This paper has reviewed some representative methods for measuring the dynamical complexity of digitized
chaotic systems from the perspective of complex networks. The general
framework of such analyses is established to facilitate the comparison of performances of comparable
analytic methods. Some subtle properties of the state-mapping network of the Logistic map operating on
a fixed-point arithmetic computer have been analyzed and discussed. This paper
demonstrates that the methodology of complex networks provides a novel tool for studying the complex
dynamics of digital chaos.

\bibliographystyle{IEEEtran_doi}
\bibliography{network_iscas19}
\end{document}